\documentclass[a4paper]{jpconf}
\usepackage{graphicx}
\begin{document}
\title{Radiative cascades in charged Quantum Dots}

\author{E Poem, Y Kodriano, C Tradonsky and D Gershoni}
\address{Department of physics, The Technion - Israel institute
of technology, Haifa, 32000, Israel}
\author{B D Gerardot and P M Petroff}
\address{Materials Department, University of California Santa
Barbara, CA, 93106, USA}
\ead{poem@technion.ac.il}

\begin{abstract}
We measured, for the first time, two photon radiative cascades due
to sequential recombination of quantum dot confined electron hole
pairs in the presence of an additional spectator charge carrier.
We identified direct, all optical cascades involving spin blockaded
intermediate states, and indirect cascades, in which non radiative
relaxation precedes the second recombination. Our
measurements provide also spin dephasing rates of confined carriers.
\end{abstract}

\begin{figure}[tbh]
\begin{minipage}[b]{15pc}
  \caption{(a) Schematic description of the energy levels of a
  singly positively charged QD.
  Vertical (curly) arrows indicate radiative (non-radiative) transitions between these levels.
  State occupation and spin wavefunctions are described to the left of each
  level where $\uparrow$ ($\Downarrow$) represents an electron (hole) with spin up (down).
  A short blue (long red) arrow represents a carrier in its first (second) level.
  S (T) stands for two holes' singlet (triplet) state and 0 (1) for $S_z=0$ ($S_z=\pm1$) total holes' pseudo-spin
  projection on the QD growth direction. The ground staste singlet is indicated by S$_G$. Only one out of two (Kramers') degenerate states is described. (b) Measured PL spectrum 
  }
\label{fig:1}
\end{minipage}
\vspace{0.25pc} \hspace{0.2pc}
\includegraphics[width=0.6\textwidth]{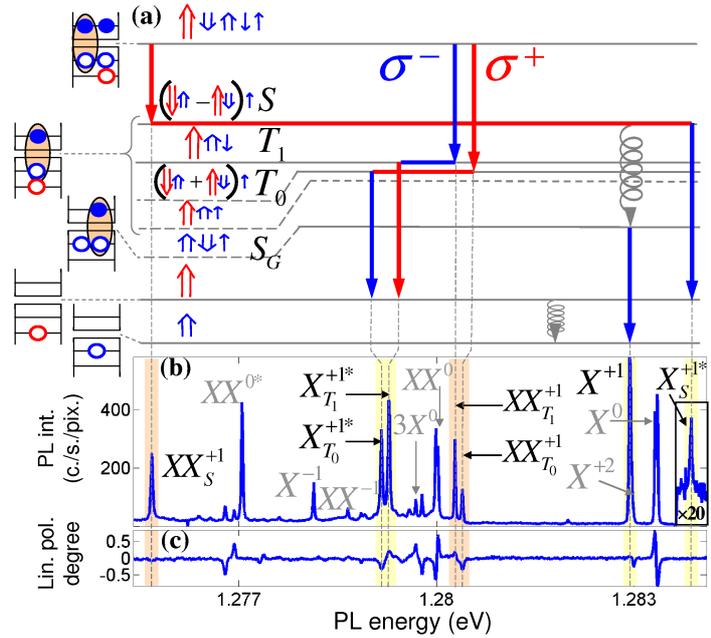}
\vspace{0.25pc}
\begin{minipage}{38pc}
on which the actual transitions are
identified. Excitonic (biexcitonic) transitions are highlighted yellow (orange).
Transitions which are not discussed here are marked by gray letters.
(c) Measured degree of linear polarization spectrum, along the
in-plane symmetry axes of the QD. Positive (negative) value
represents polarization along the QD's major (minor) axis.
\end{minipage}
\end{figure}

Semiconductor quantum dots (QDs) strongly localize charge carriers,
and discretize their energy level spectrum, in a similar way to
electrons in atoms. 
Radiative cascades in neutral
QDs~\cite{Gerard_PRL01,Regelman_PRL,Kiraz_PRB,Santori_PRB02}
demonstrated their potential as deterministic sources for
polarization entangled photon pairs~\cite{nika_ent,michlerNJP}. The
neutral radiative cascade~\cite{gammon,kulakovskii} leaves the QD
empty of charge carriers. This is essential for entangling the
emitted two photons, since otherwise the remaining carrier's spin
betrays the required ``which path"
ambiguity~\cite{Tejedor,avron_time_reord}. Neutrality prevents,
however, the important benefit of correlating between the emitted
photons' polarizations (``flying qubits") and the local carrier's
spins (``anchored qubits"). The situation is drastically different
in charged QDs, where quantum correlations exist between the flying
and anchored qubits.
Here we report on two-photon radiative cascades in the
presence of an additional hole.\\ 
The energy levels of a positively charged QD~\cite{kavokin,akimov}
containing up to three holes and two electrons are schematically
described in \mbox{Fig. \ref{fig:1}(a)}. The figure presents also
the relevant optical and non-optical total-spin conserving
transitions between these levels. The two photon radiative cascades
start from the ground level of the three hole and two electron
state. 
The unpaired hole's spin projection
along the growth axis determines the total spin of the two Kramers'
degenerate states (for simplicity only one state is drawn in
\mbox{Fig. \ref{fig:1}(a)}). Radiative recombination of first level
electron-hole (e-h) pair leaves three unpaired charge carriers
within the QD. There are 8 possible different spin configurations
for the remaining carriers. These configurations form 4 energy
levels of Kramers' pairs
~\cite{kavokin,akimov}. The three lowest levels are those in which
the two unpaired holes are in spin-triplet states. Those states are
separated from the highest energy level in which the holes are in a
singlet spin state by the hole-hole isotropic exchange interaction,
which is significantly stronger than the e-h exchange interaction.
The later removes the degeneracy between the triplet states as shown
in \mbox{Fig. \ref{fig:1}(a)}. The lowest triplet level
cannot be reached optically. The
optical transitions into the
other levels are optically allowed.
The circular polarization of the emitted photons are indicated in
the figure. It depends on the spins of the annihilated electron hole
pair.
The measured emission contains also linear components (see
\mbox{Fig. \ref{fig:1}(c)}), due to the anisotropic e-h exchange
interaction~\cite{kavokin,akimov}.
The relaxation proceeds by radiative recombination of the remaining
first level e-h pair, leaving thus only one hole in its second
level. The hole can then quickly relax non-radiatively to its ground
level. There is a fundamental difference between the singlet and
triplet intermediate states. While in the later, due to Pauli's
exclusion principle, radiative recombination must occur before the
excited hole can relax to its ground state (resulting in two
``direct" cascades), in the former non-radiative relaxation of the
excited hole state may occur prior to the radiative recombination
(resulting in one ``direct" and one
``indirect" cascade).

\begin{figure}[tbh]
\begin{minipage}{18.5pc}
  \includegraphics[width=1\textwidth]{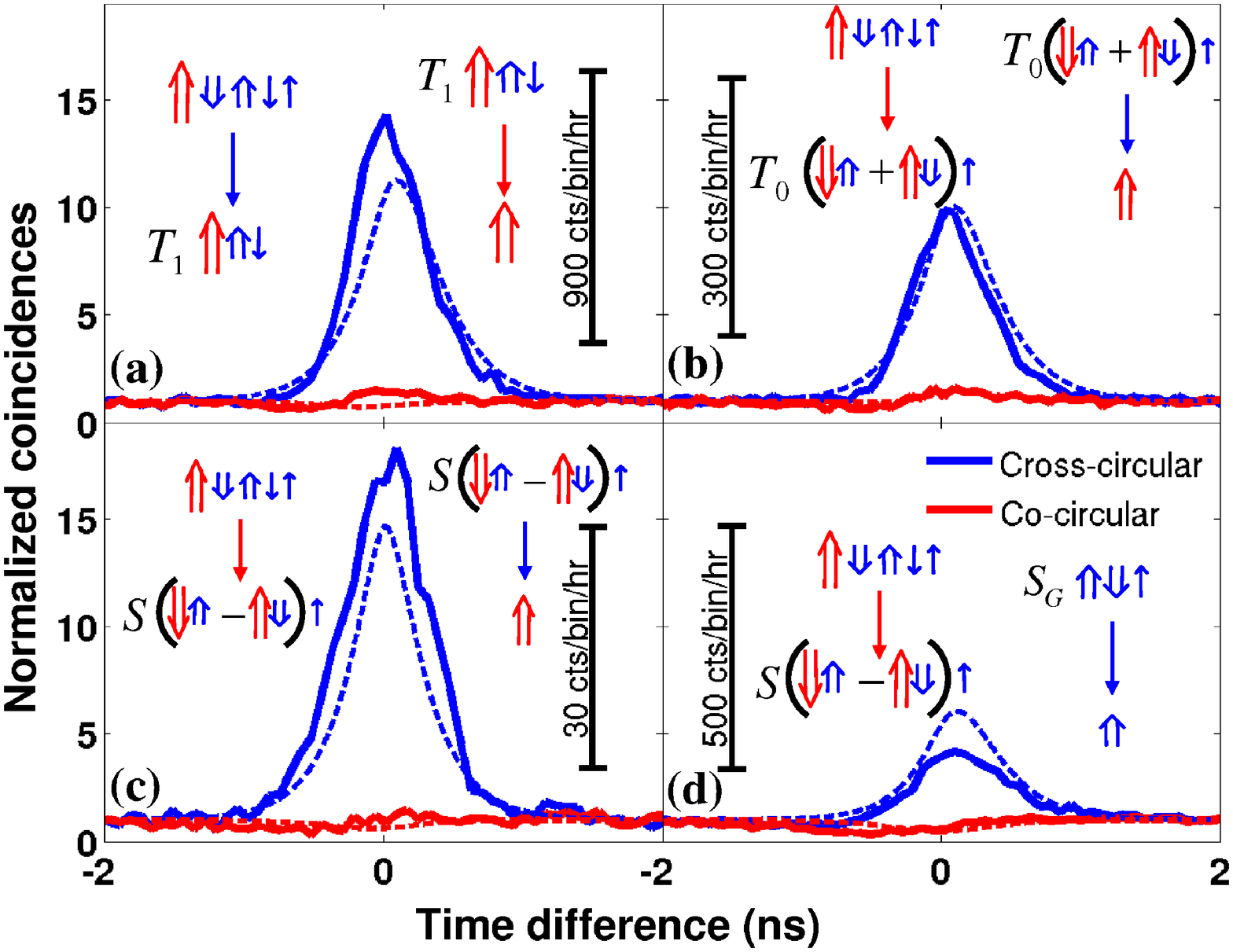}
  \caption{Measured and calculated time-resolved, polarization sensitive
  intensity correlation functions, for the 4 radiative cascades described in
\mbox{Fig. \ref{fig:1}}.
  The states involved in the first (second) photon emission are illustrated to the left
  (right) side of each panel. All
   symbols and labels are as in \mbox{Fig. \ref{fig:1}}.
  Solid Blue (red) line stands for measured cross- (co-)
  circularly polarized photons. Dashed lines represent the corresponding
  calculated functions.
  The bar presents the acquisition rate in coincidences per time
  bin (80 ps) per hour.}
  \label{fig:2}
\end{minipage}\hspace{1pc}%
\begin{minipage}{18.5pc}
  \includegraphics[width=1\textwidth]{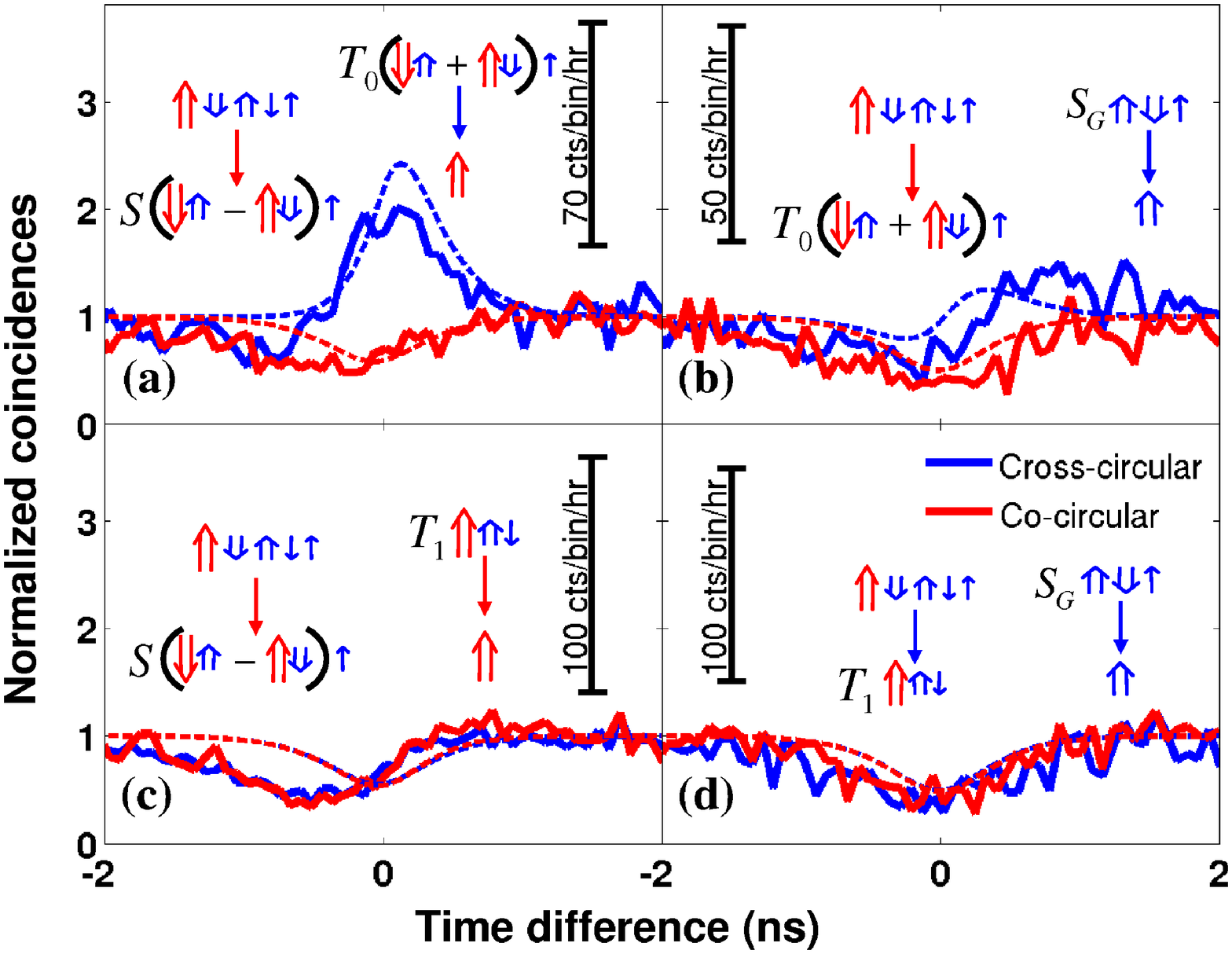}
  \caption{Measured and calculated time-resolved, polarization sensitive
  intensity correlation functions, across the radiative cascades.
  (a) [(c)] Correlations between the singlet biexciton transition and the exciton transition from the
  $T_0$, [$T_1$] state. (b) [(d)] Correlations between the $T_0$, [$T_1$] biexciton transition
  and the ground $X^{+1}$ exciton transition.
  All symbols and labels are as in \mbox{Fig. \ref{fig:1}}. The meanings of all line types and colors are as in \mbox{Fig. \ref{fig:2}}.
  }
\label{fig:3}
\end{minipage}
\end{figure}

The studied sample contains InGaAs QDs in the middle of a 1$\lambda$
microcavity \cite{nika_ent}.
For the optical measurements the sample was placed inside a tube
immersed in liquid Helium, maintaining sample temperature of 4.2K. A
X60, 0.85 NA, in-situ microscope objective was used to both focus
the exciting beam on the sample surface and to collect the emitted
light.
The polarization of the emitted light was analyzed using two
computer controlled liquid crystal variable retarders and a linear
polarizer. In \mbox{Fig. \ref{fig:1}(b)} we present the spectrum
measured under non-resonant cw excitation with 1~$\mu$W of HeNe
laser light (1.96 eV). The corresponding degree of linear
polarization is presented in \mbox{Fig. \ref{fig:1}(c)}. The
spectral lines participating in the radiative cascades described in
\mbox{Fig. \ref{fig:1}(a)} are clearly identified spectrally in the
single QD PL and linear polarization spectra, and are highlighted
orange (yellow) for biexcitonic (excitonic) transitions. For
polarization-sensitive time-resolved intensity-correlation
measurements, we used a Hanbury-Brown and Twiss like apparatus
~\cite{nika_ent}.
In \mbox{Fig. \ref{fig:2}} we present the measured and calculated
intensity correlation functions for photon pairs emitted in the four
spin-conserving radiative cascades outlined in \mbox{Fig.
\ref{fig:1}(a)}. 
The measured data clearly reveal the sequence of the radiative
events, reassuring the interpretations of \mbox{Fig. \ref{fig:1}}.
In \mbox{Fig. \ref{fig:3}} we present measured and calculated
intensity correlation functions {\bf between} different radiative
cascades.
Since spin blockading prevents the relaxation of the second level
hole to its first level, they provide an estimate for the rate by
which the hole spin's scatters~\cite{Eilonssc}. Fast scattering would give rise to a peak in the correlation function, because then the photon emissions preceeding and succeeding the scattering process would mostly happen one right after the other. Scattering rate slower than the radiative recombination rate and/or the optical generation rate would give rise to a dip in the correlation function, since the second photon would most probably be emitted only after additional recombination and generation of e-h pairs.
In \mbox{Fig. \ref{fig:3}} (a) and (c) we probe possible transitions
from the singlet intermediate state to the triplet $T_0$ and $T_1$
intermediate states, respectively. In (b) and (d) we probe possible
transitions from the triplet $T_0$ and $T_1$ intermediate states,
respectively, to the singlet ground state. Assuming that relaxation
from the intermediate triplet states to the ground singlet states
must be preceded by transition to the intermediate singlet states,
these measurements provide quantitative estimation for the reverse
of the processes described in (a) and (c). From the measured data in
\mbox{Fig. \ref{fig:3}} one clearly notes that transition between
the two holes' singlet state to the $T_1$ triplet state (\mbox{Fig.
\ref{fig:3}c}) and vice versa (\mbox{Fig. \ref{fig:3}d}) are
forbidden,
while
transitions between the singlet and the $T_0$ triplet states
(\mbox{Fig. \ref{fig:3}a}) and vice versa (\mbox{Fig. \ref{fig:3}b})
are partially allowed. This means that the holes spin projection on
the QD's growth axis is conserved during the relaxation while their
in-plane spin projection scatters~\cite{Eilonssc}. The difference
between the scattering rates from the singlet to triplet state and
that from the triplet to singlet is due to the energy difference
between these two states ($\sim$4 meV), which is much larger than
the ambient thermal energy ($\sim$0.5 meV).\\
Our model is composed of a set of coupled rate equations for the
time-dependent probabilities of finding the system in one of its
many-carriers-states~\cite{Regelman_PRL}.
We include all the states as described in
\mbox{Fig. \ref{fig:1}(a)}, together with their Kramers conjugates.
In addition, we include four more states representing charged
multiexcitons up to 6 e-h pairs~\cite{Regelman_PRL}. There are clear
spectral evidences for processes in which the QD changes its charge
state and becomes neutral due to optical
depletion~\cite{hartmann,baier,Michler_PRB05} (see \mbox{Fig.
\ref{fig:1}}(b)). These observations are considered in our model by
introducing one additional state which represents a neutral QD. The
transition rates between the states include radiative rate ($\gamma_r
= 1.25 ns^{-1}$ deduced directly from the PL decay of the exciton
lines) and non-radiative spin-conserving rate ($\Gamma_{S
\rightarrow S}=35\gamma_r$, deduced from the intensity ratios of the
relevant PL lines).
We also include the rates for optical generation of e-h pairs
($G_e=1\gamma_r$, forced by equating between the emission
intensities of the biexciton and exciton spectral lines), optical
depletion, and recharging ($G_D=4\gamma_r$ and $G_C=0.1\gamma_r$ as
deduced from the relevant line intensity ratios, and correlation
measurements between the neutral and charged exciton).

The data clearly show that hole spin scattering rates, ($\Gamma_{S
\leftrightarrow T_1}$ ) which do not conserve the spin projection on
the QD's growth axis, are vanishingly small. Therefore we set them
to 0. In order to account for the observed correlations between
singlet ($S$) and $T_0$ states, \mbox{Fig. \ref{fig:3}}, we fitted
in-plane scattering rates~\cite{Eilonssc}
$\Gamma_{T_0 \rightarrow S}= 0.6 \gamma_r$ and $\Gamma_{S
\rightarrow T_0} = 10 \gamma_r$ (such processes still conserve the
projection of the total spin along the growth axis). The ratio
between these rates simply
gives the temperature of the optically excited QD ($\sim$19K).
The anisotropic e-h exchange interaction mixes the $T_0$ and $T_1$
states~\cite{akimov,kavokin}. This makes the natural polarizations
of the relevant transitions elliptical rather than circular. The
mixing degree is obtained from the measured degree of linear
polarization of the biexciton transitions~\cite{akimov}. Our model
considers this mixing as well.
It explains the non-zero correlations in co-circular polarizations.

The 2 inplane hole's spin scattering rates that we fitted describe
very well the 16 measured intensity correlation functions. The
calculated functions convoluted with the system response are
presented in \mbox{Fig. \ref{fig:2}} and \mbox{Fig. \ref{fig:3}} by
dashed lines.

In summary, we identified 3 direct and one indirect radiative
cascades in singly charged QDs and demonstrated unambiguous
correlations between the polarizations of the emitted photons and
the spin of the remaining charge carrier. Our correlation
measurements show that while holes' spin-projection conserving
scattering rates are few times faster than the radiative rates,
spin-projection non-conserving rates are vanishingly small.

\ack The support of the US-Israel binational science foundation
(BSF), the Israeli science foundation (ISF), the Israeli ministry of science
and technology and the Technion's RBNI are gratefully acknowledged.

\section*{References}
\bibliographystyle{iopart-num}
\bibliography{pos_casc_full}

\end{document}